\documentclass[11pt]{article}
\usepackage{graphicx} 
\usepackage{caption}
\usepackage{amsmath}
\usepackage[T1]{fontenc}
\usepackage[utf8]{inputenc}
\usepackage{authblk}
\usepackage{hyperref}

\title{From learning gait signatures of many individuals to reconstructing gait dynamics of one single individual.}
\author[1]{Fushing Hsieh}
\author[1]{Xiaodong Wang}
\affil[1]{Department of Statistics, University of California, Davis.}
\date{} 

\begin{document}

\maketitle

\begin{abstract}
Based on the same databases, we computationally address two seemingly highly related, in fact drastically distinct, questions via computational data-driven algorithms: 1) how to precisely achieve the big task of differentiating gait signatures of many individuals? 2) how to reconstruct an individual’s complex gait dynamics in full? Our brains can ``effortlessly’’ resolve the first question, but will definitely fail in the second one. Since many fine temporal scale gait patterns surely escape our eyes. Based on accelerometers' 3D gait time series databases, we link the answers toward both questions via multiscale structural dependency within gait dynamics of our musculoskeletal system. Two types of dependency manifestations are explored. We first develop simple algorithmic computing called Principle System-State Analysis (PSSA) for the coarse dependency in implicit forms. PSSA is shown to be able to efficiently classifying among many subjects. We then develop a multiscale Local-1st-Global-2nd (L1G2) Coding Algorithm and a landmark computing algorithm. With both algorithms, we can precisely dissect rhythmic gait cycles, and then decompose each cycle into a series of cyclic gait phases. With proper color-coding and stacking, we reconstruct and represent an individual’s gait dynamics via a 3D cylinder to collectively reveal universal deterministic and stochastic structural patterns on centisecond (10 milliseconds) scale across all rhythmic cycles. This 3D cylinder can serve as ``passtensor'' for authentication purposes related to clinical diagnoses and cybersecurity.
\end{abstract}

\section{Introduction}
It seems ordinary that we recognize our close friends and family members by their distinctive walking "styles", so-called signatures of gaits. With the complexity of neural and musculoskeletal systems in mind \cite{Winter}, the gait dynamics is not at all simple. Unlike high speed camera, our eyes surely miss all gait patterns of fine temporal scales. So, this ability of ours is not at all ordinary. Even though we human are anatomically identical by sharing the same structural skeleton and muscle constructs, and any gait dynamics must obey the universal biomechanics governing our musculoskeletal system, what make up individual signatures of gaits as biometric traits is still not yet well understood.

Majority of gait related research works is in the category of modeling-based gait analyses. The whole gait dynamics is never the focus. Any model based on only a few characteristics of gait dynamics typically not only is prone to make mistakes, but also difficult to apply to large number of healthy people. For instance, many works mainly aim for either Parkinson disease predictions or risk evaluations for the elderly \cite{Ailisto, Gafurov, Lai, Trivino, Whittle}. Such top-down approaches are of limited used for surveillance, for example, because they don’t embrace diverse spectra of gait characteristics. For instance, the fuzzy finite state machine \cite{Alvarez} needs to incorporate expert opinions and judgements for specifying relevant states. Further transitions between states are governed by fuzzy logics \cite{Zadeh}.

Recently data collecting technologies have drastically evolved with recent advances in Microelectromechanical systems (MEMS), such as low-cost, light-weight, easy-to-use inertial measurement units (IMU), such as accelerometer and gyroscope sensors \cite{Sprager}. These sensors nowadays are integrated with mobile devices, which enable us to collect gait time series data outside of gait laboratory, see figures of human wearing sensors in \cite{Khandelwal, Ngo}. However, the capacity of precisely differentiating many subjects’ gait signatures and seeing a person's multiscale gait dynamics in full are not yet available in literature.



\begin{figure*}[h]
\centering
\includegraphics[width=5.3in]{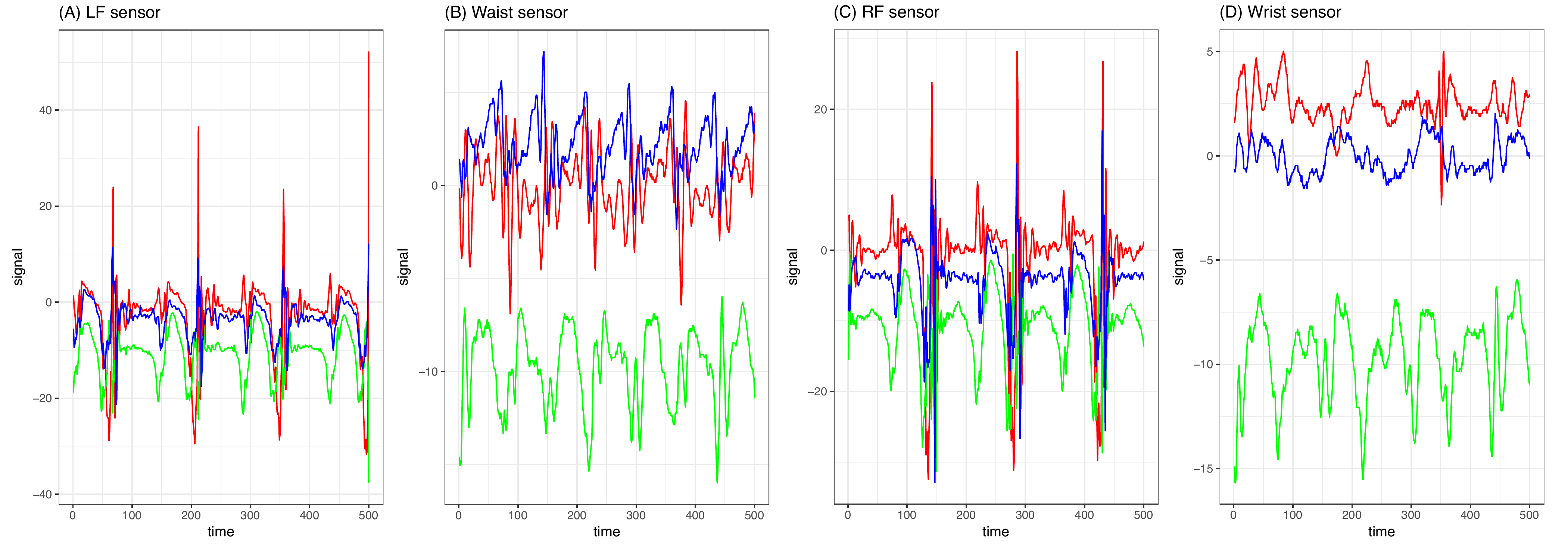}
\caption{Gait time series data of subject $\# 5$ from four sensors: (A)Left foot; (B)Waist; (C)Right foot; (D)Wrist. $X-$dimension is Red color-coded, $Y-$dimension is Green and $Z-$dimension is Blue.}
\label{fig.gaitdata}
\end{figure*}



In this paper, we develop computing and data-driven algorithms suitable for addressing two questions. 1) How to find and embrace large and diverse spectra of gait characteristics for identification purpose; 2) How to discover and recreate a person's gait dynamics in full? 

The first theme of our data-driven developments is to compute and find many principle directions or vectors that implicitly capture many important aspects of above structural dependency-based heterogeneity across many people. We consider one manifestation of structural dependency through temporal patterns via a very simple and coarse coding scheme, called Principle System-State Analysis (PSSA). This dependency manifestation of coarse scale pattern is indeed very versatile for classifying among all subjects. We conjecture that this kind of dependency manifestation is potentially close to how our brains learn gait signatures.

As a complex system, and the intelligence of musculoskeletal system is embraced by its multiscale heterogeneity \cite{Anderson}. It is well known that any real ``rhythmic'' biomechanics is far from being completely deterministic and it naturally embraces stochastic structures across all rhythmic cycles as well \cite{Crutchfield}. Here it is worth emphasizing the evidently visible, but inexplicable stochasticity. Since this stochasticity is chiefly constrained by deterministic structures, it is not completely random. Therefore extracting stochastic structures of gait dynamics is at least as equally important as extracting the deterministic counterparts.

For explicitly extracting such multiscale deterministic and stochastic information contents, we turn to and focus on the system's fundamental structural dependency among all observed gait time series. It is clear that such structural dependency is lost to a great degree in the so-called resultant acceleration signal \cite{Santanna, Karantonis}:
\[
A_{res}[t]=\sqrt{X^2[t]+Y^2[t]+Z^2[t]},
\]
This fact is evident through our motivating Lampel-Ziv complexity experiments, see details in the next section. Results from such experiments imply how to build a symbolic coding scheme to retain structural dependency of multiple time series.

Based on such motivation, our second theme of data-driven computing paradigm is developed as an unsupervised learning based multi-layer coding scheme, called Local-first and Global-second (L1G2) coding scheme. We apply L1G2 to build a 2D code sequence pertaining to the [Left-foot + Right-foot] system. We also develop a landmark partition algorithm to dissect such a 2D code sequence into rhythmic cycles consisting of visible biomechanical states. Such rhythmic patterns confirm that this subsystem indeed dictates the contents of a rhythmic cycle, its period and most importantly its evolving process. That is, the entire musculoskeletal system should function by coupling others subsystems upon [Left-foot + Right-foot] system.

To further show L1G2 effectively capturing multiscale gait dynamics, via graphic display, we simply stack all resultant color-coded rhythmic cycles aligned with the landmarks into a 3D cylinder. This rotatable 3D cylinder coherently reveals multiscale deterministic and stochastic rhythmic patterns as multiscale structural dependency across all rhythmic cycles. Such a 3D cylinder is the very foundation of further researches of gait-mimicking. It is also good for clinical diagnosis, and can be used as a ``passtensor'' for cybersecurity.

Two known gait time series databases are analyzed as the real data experiments. 1) MAREA database \cite{Khandelwal} with 4 sensors; 2) HuGaDB database \cite{Chereshnev} with 6 sensors. Both databases are created on healthy subject's gait when subjects wear with multiple sensors performing various activities on different kinds of surfaces. The sampling rate in MAREA is 128Hz, and is less in HuGaDB. That is, the time series in these databases contains patterns of centisecond (10 mini-second) scale.

We focus only on accelerometer in this paper. It picks up accelerations of linear motions of body parts, where the sensors are fixed, upon $X-$, $Y-$ and $Z-$axial orientations. The 3-dim measurements are referencing to the coordinate system of human body: anterior-posterior (forward vs backward), superior-inferior (vertical up vs down along gravity direction) and left-right \cite{Gietzelt}. Our developments can easily accommodate gyroscope-based time series.

\section{Revelations of structural dependency}
To set the stage for our computational developments for exploring an individual's gait dynamics in full, we give an overview of the two contrasting manifestations of structural dependency contained in multi-dimensional gait time series. First by looking at an approximate 3 sec. recording of 12 dimensional time series of a MEARA subject's walking on indoor flat ground, as shown in Fig.\ref{fig.gaitdata}, we see that each sensor's triplet directional time series exhibit diverse scales of relational patterns, which evolve within each visible cycle, and recurrently appear across evident rhythmic cycles. Secondly, when we compare such patterns across different sensors, we also discover various scales of recurrent pattern-to-pattern correspondences. Such pattern-to-pattern correspondences are especially evident between panel (A) of Left-foot and panel (C) of Right-foot of Fig.\ref{fig.gaitdata} across the evident cycles. Pattern-to-pattern correspondences between panel (B) of Waist and either one of Left-foot or Right-foot are also apparent, but not between panel (D) of Wrist with the rest of panels. These visible temporal-oriented relational patterns within cycles and complex pattern-to-pattern correspondences across cycles constitute multiscale structural dependency of gait dynamics contained in the 12 dimensional time series. This is the chief concern in this paper.

\begin{figure*}[h]
\centering
\includegraphics[width=5.2in]{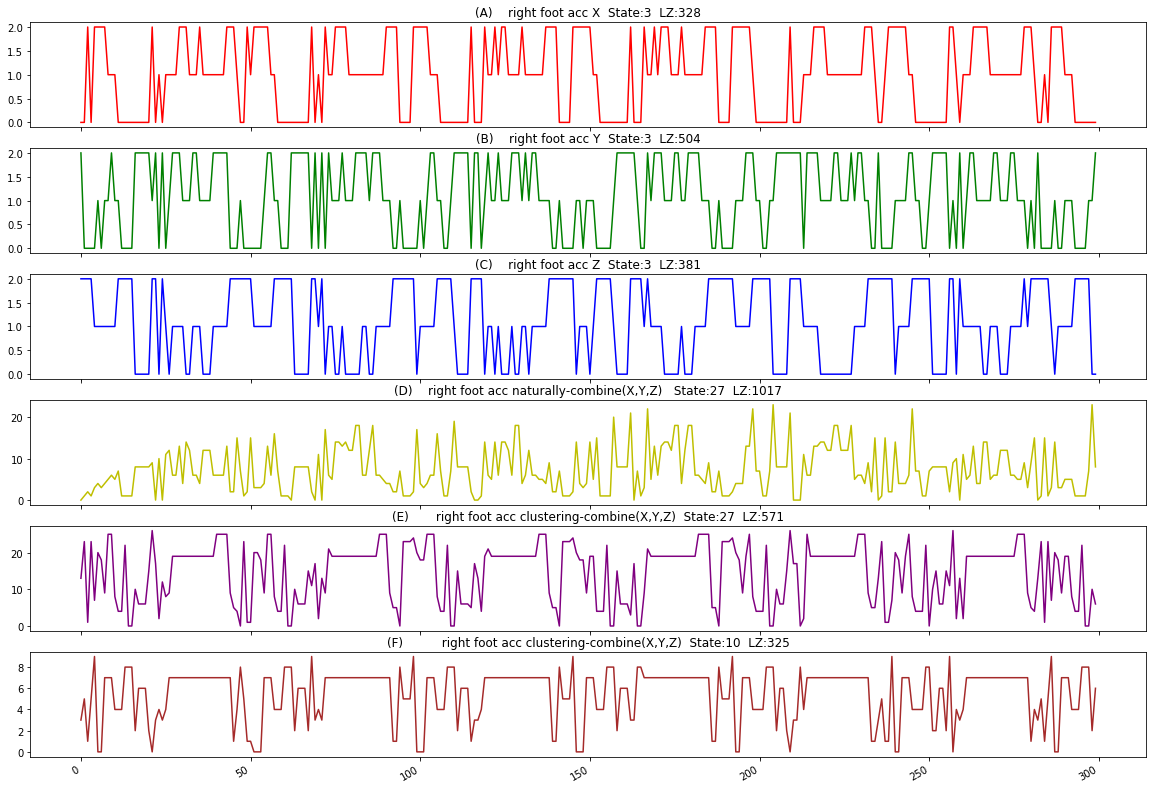}
\caption{(A),(B),(C) 3-state code sequences for X-,Y-,Z- accelerometer time series, respectively. (D) is a natural combination of X,Y,Z, and the resultant sequence is coded by 27 ($3\times3\times3$) states. (E),(F) are sequences based on our clustering-based way of combination; (E) is coded by 27 states (clusters), the same number of states as (D), while its LZ complexity reduces by half. (F) a 10-states code sequence can show the rhythmic pattern clear enough, and its LZ complexity is as low as that of one-dim time series case.}
\label{fig:lzcomplexity}
\end{figure*}

In computational theory of computer science, the concept of Kolmogorov complexity is used in evaluating and exploring hidden structural patterns embraced within symbolic or digital time series. Its conceptual shortest universal computer program for regenerating a time series at hand is recognized to embrace all deterministic and stochastic structures. Unfortunately, Kolmogorov complexity cannot be calculated in general. We employ Lempel-Ziv complexity to give an approximate measure by only using ‘copy’ and ‘insert’ two operations. This complexity can be efficiently computed, see \cite{Kaspar}. So, Lempel-Ziv is used in our complexity experiments. Before our complexity experiment, all the continuous time series must be categorized and transformed into a finite and discrete state sequence. 

As shown in each panel of Fig.\ref{fig.gaitdata}, each triplet time series of $(X, Y, Z)$ directions of an accelerometer reveal varying mechanism-specific gait dynamic patterns. Thus, we make use of this data transformation requirement to naturally link the concept of structural dependency among time series to system-states of its dynamics. The idea of system-state can be seen as follows. We develop two tempo-sensitive digital-coding schemes upon gait time series along the temporal axis. The first scheme is to perform digital-coding upon each of the triplet directional time series individually and then couple the three digital code sequences into one sequence of vectors. The second scheme is to apply Hierarchical clustering algorithm on the temporal (column) axis of a data matrix representing the triplet time series with 3 rows. Based on the resultant clustering tree, a composition of clusters is chosen. A cluster of 3D vectors can be regarded as a symbolic code for a system state. Hence the specific mechanism pertaining to an accelerometer along the temporal axis is represented by a 1D symbolic code sequence. Color-coded examples of such code sequences are given in Fig.\ref{fig:coding}. The computing cost of the first approach is much less than that of second approach. But, unlike the second approach, the first approach can only capture relatively coarse structural dependency. 

We compare these two coding schemes in a set of Lampel-Ziv complexity experiments based on a short temporal segment $[0, 300]$. Results of such experiments are summarized in Fig.\ref{fig:lzcomplexity}, also see Fig.\ref{fig:compare_rf}, \ref{fig:compare_lf} in Supplementary for more details. The top three panels of Fig.\ref{fig:lzcomplexity} respectively give the three directional symbolic code sequences. Each code sequence has 3 states and a value of Lampel-Ziv complexity. By coupling these three code sequences along the temporal axis, as shown in panel (D), the resultant code sequence with 27 state is seen nearly without any recognizable recurrent patterns. It has a complexity value 1017. In comparison, the second scheme with 27 clusters results into code sequence, as shown in the panel (E), that shows very evident recurrent and rhythmic patterns with a complexity value 571. Further, even if only 10 clusters are used to form the set of states, as seen in the bottom panel (F), the resultant code sequence is as evidently rhythmic as the one with 27 states in (E). With such rhythmic patterns in view, it is not surprising that its Lampel-Ziv complexity value is even lower. Evidently it captures the rhythmic dynamics well. Such experimental results confirm the presence of structural dependency among the three directional gait time series, and at the same time imply that the second coding scheme is way of extracting detailed dependency patterns in gait dynamics. Nonetheless, the first coding scheme has its own merit in identifying among many subjects as seen in the next section.

\section{Principle System-State Algorithm (PSSA) for Identification.}
A simple way of having a glimpse of structural dependency among sensor-direction specific $D$ dimensional gait time series is to transform and couple them into a $D$-dimensional digital vector trajectory. Here $D$ is equal to 12 for 4 sensors used in MAREA database and 18 in for 6 sensors used in HuGaDB database. This digital trajectory is to exhibit rough manifestations of rhythmic cycles. So we manage to have a representation with relative small algorithmic complexity about the gait dynamics. This idea is simple and intuitive. Here we develop data-driven computations via a coarse coding scheme to realize this concept. By doing so, we get away from the necessity of man-made system-states and requirements of their transition rules. The simple computational results are capable of identifying many subjects simultaneously on a single platform.  Thus we speculate such a simple algorithm is potentially what our brain actually performs in recognizing friends and relatives' gait signatures.
To this aim, we propose an algorithm, called the Principle System-State Algorithm (PSSA), that attempts a single-layer coarse structural dependency among many individuals' $D$ dimensional gait time series simultaneously.

\subsection{PSSA algorithm}
For the purpose of identification, we expect to identify an individual by only glimpsing his/her short time of walking. Each individual's specific gait time series is subdivided into replicates of period in equal length $l$. we assume that in the test set, each unlabeled individual would have sample size exceeding $l$. The choice of $l$ is supposed to be small while the signal is strong enough. Here we set $l=1000$ time points, which lasts about 7.7 seconds with respect to the sampling rate being set at 128Hz. The PSSA algorithm is described below.

First, encode each sensor-direction specific 1-dim time series by using 3-digit alphabets.
\[ S_d(t) = \begin{cases} 1 & \quad X_d(t)\le\alpha\\
2 & \quad \alpha < X_d(t)\le\beta\\ 
3 & \quad X_d(t)>\beta\\ 
\end{cases} \]
where $X(t)$ is the variable at time stamp $t$ and $d=1,2,...,D$ indicating dimension. So a $D$-dimensional digital system-state (vector), say $S(t)=(S_1(t), ...., S_D(t))^{'}$, is formed at each time point $t$. The tuning parameter $\alpha$ and $\beta$ ($\alpha < 0.5 < \beta$) are chosen based on the quantile of each $1$-dim empirical distribution of pooled data across all involving subjects. The complexity of resultant digital code time series becomes smaller if both $\alpha$ and $\beta$ are closer to their extremes 0 and 1, respectively.

Second, collect all distinct system-states $S(.)$ and calculate their corresponding frequency $f$. There will be at most $3^D$ possibilities. Sort the distinct states with respect to frequency from the most frequent to the least
$(S^{(1)}(.),..., S^{(N)}(.))^\prime$ with highest frequency $f^{(1)}$ to the lowest one $f^{(N)}$. Select a set of $N^*$ states with top highest frequency as principle system-states (PSS).

Third, cut Gait time series from the training set into short-temporal segments in length $l$, and convert each segment to a $N^*$-vector of proportion of PSS occurring within the period.

Finally, build a $m\times N^*$ rectangle matrix $\Sigma_{PSS}$ by stacking all involving proportion vectors along the row-axis, where $m$ is the total number of segments. Apply hierarchical clustering analysis on row and column axes of $\Sigma_{PSS}$, respectively. Find the corresponding `key' PSS for each individual such that the PSS can be used as a new feature (group) to exclusively identify the individual from others.

PSSA achieves a huge reduction on temporal dimensionality from $l=1000$ to $N^*$. More importantly, such a $N^*$-dim vector is in the category of structural data, that is, each component can be treated as a feature variable. So any classic machine learning techniques can come in and work on the structured matrix $\Sigma_{PSS}$.

With a chosen pair of tuning parameter $\alpha$ and $\beta$ ($\alpha < 0.5 < \beta$). the complexity digital coded $D$-dim time series can be seen via the curve of proportion of coverage on all involving trajectories as:
\[
r(N^*)=\sum^{N^*}_{i=1} f^{(i)}/N,
\]
The selection of $N^*$ principle system-states $(S^{(1)}(.),..., S^{(N^*)}(.))$ can be also based on this curve.

\subsection{PSSA on real databases}
Two examples of coverage proportion curves with respect to $N^*$ principle system-states are given Fig.\ref{fig:PSS_CurveA},\ref{fig:PSS_CurveB} in Supplementary for MAREA database and HUGaDB database.

\begin{figure*}[h]
 \centering
 \includegraphics[width=5in]{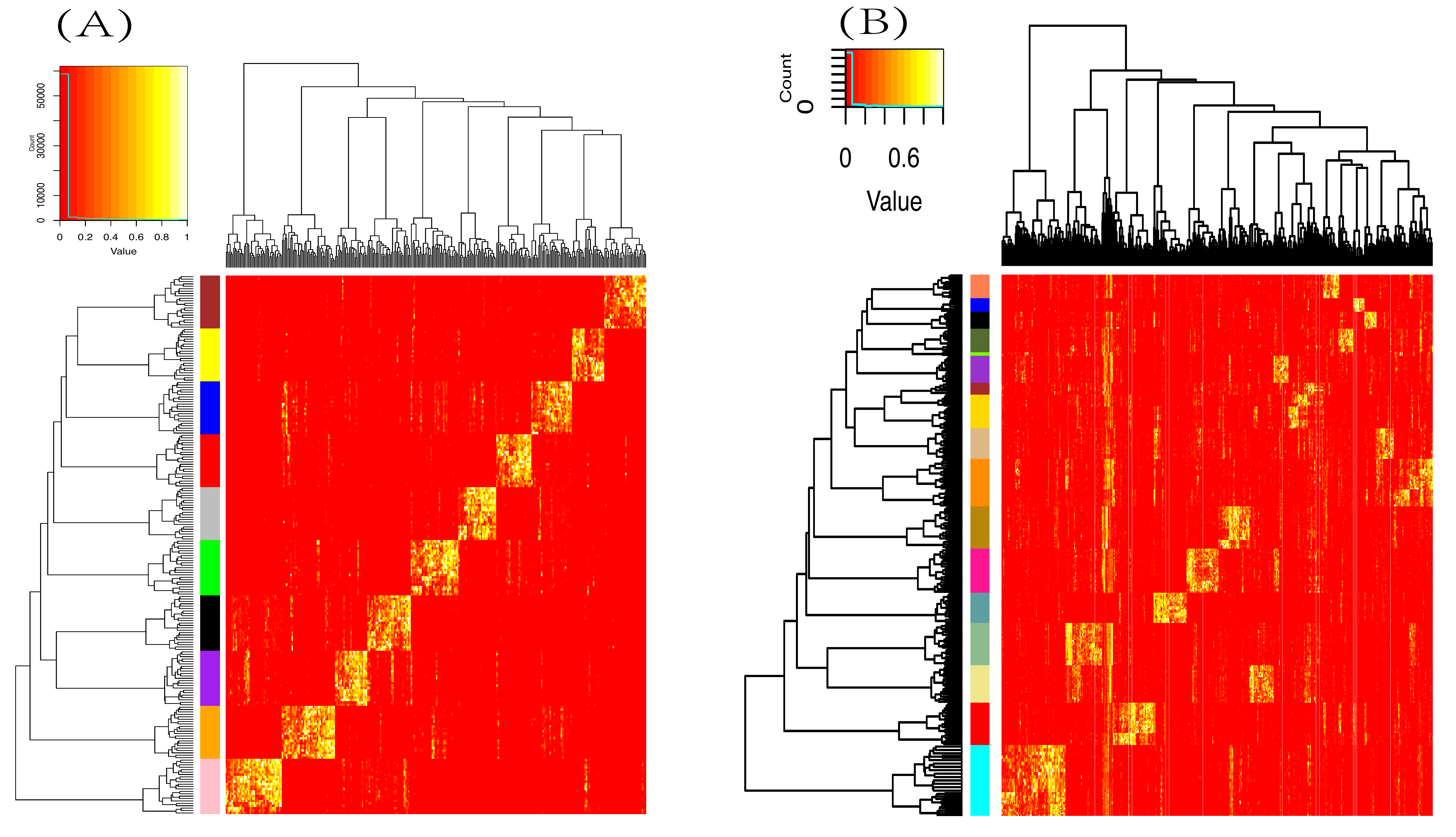}
 \caption{Identification via heatmap of $\Sigma_{PSS}$. Each row indicates a segment of gait time and rows from the same subject are labeled in the same color; each column indicates a selected PSS (A) MAREA database: 10 subjects. The quantiles $\alpha=0.3$ and $\beta=0.7$. $N^*(=300)$ principle system-states based on 9 dimensions of gait time series derived from three sensors fixed at Left foot and Right foot and wrist; (B)HuGaDB database: 17 subjects with 6 sensors tied to left and right thighs, shins and feet. The quantiles $\alpha=0.1$ and $\beta=0.9$. $N^*(=500)$ principle system-states based on 18 dimensions of gait time series.}
 \label{fig:gaitid}
 \end{figure*}

Both results in the training set are perfect classified without any error among all 10 subjects' replicates in MAREA database, and 17 subjects' replicates in HuGaDB database, see Fig.\ref{fig:gaitid}. By selecting one significant states block or cluster for each individual, a simple decision tree can achieve perfect classification result in the test set. That is to say, the principle states take the shape of feature selection, and they are the keys in Gait identification.


Here we make a remark on how to scale a big ensemble of individuals via PSSA. When the ensemble of individuals is big in size, the PSSA needs a strategy to scale down the computing loading. That is, if such an ensemble is taken as being homogeneous, then PSSA will need a large collection of system-state vectors to cover enough complexity in identification task. Or the percentages $\alpha$ and $\beta$ are chosen to be close their extremes. On the other hand, if heterogeneity is naturally present in any human ensemble, it implies the necessity of partitioning the whole ensemble into homogeneous sub-ensembles, and then PSSA is applied respectively. This is a typical divide-and-conquer strategy. For instance, the database in Ngo et al.(2014) consists of more than 700 individuals. It is sensible to divide the whole ensemble with respect to available demographic information.

In summary, our PSSA algorithm apparently is able to identify a set of system-states as signatures for each individual subject via relatively easy computations, and then perfectly classify among these subjects. Such visible signatures are indeed between-subject characteristics in nature. Since the computing behind such signatures is so simple, it is postulated why our brain can capture such signatures seemingly with easy after lengthy observations.

\section{Authentication via structural dependency}
Here if we agree that different sets of triplet time series from different sensors give rise to different aspects of gait dynamics pertaining to our musculoskeletal system, then to authentically recreate gait dynamics is equivalently to compute {\bf the multiscale structural dependency based on all available time series data}.

Let the local scale refer to various body components of musculoskeletal system, such as Left-foot, Right-foot, Waist and Wrist. Each component contributes a fixed series of nearly deterministic biomechanical phases. Each biomechanical phase involves with a specific type of stochasticity: either in lengths or compositional contents. It is worth noting that such stochastic structures are somehow constrained by deterministic structures.

Let the global scale refer to how different components of musculoskeletal system couple and work out gait dynamics. Due to their dual symmetry, we particularly focus on how Left-foot relationally works with Right-foot via an evolving process. The [Left-foot + Right-foot] subsystem is rather distinct from their relations to Waist as the center of mass with the musculoskeletal system. That is, within the entire musculoskeletal system, the [Left-foot + Right-foot] system indeed functionally coordinates with different subsystems.

\begin{figure}[h]
\centering
\includegraphics[width=3.2in]{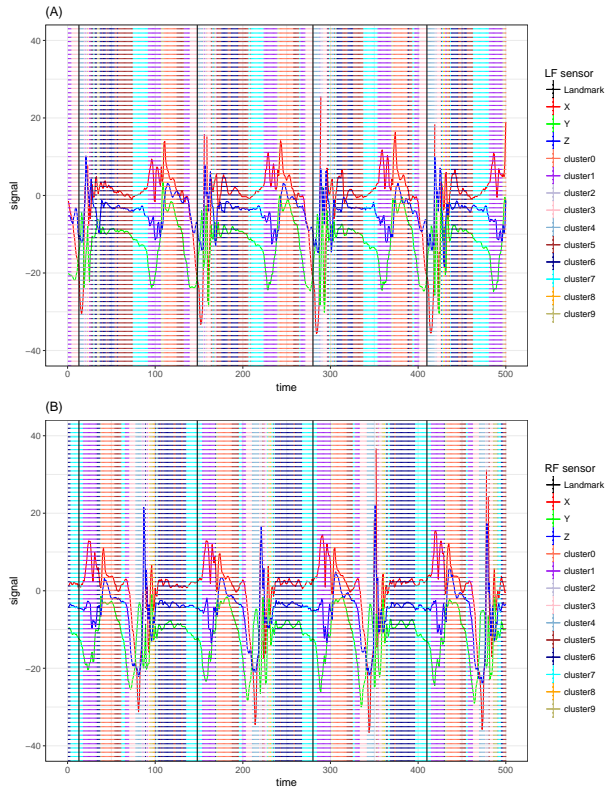}
\caption{3D time series superimposed with color coding on temporal period [1, 500]: (A) Left-foot sensor; (B) Right-foot sensor. Color coding of the 10 selected clusters are listed on the right hand side. The landmarks are calculated and marked with vertical black line.}
\label{fig:coding}
\end{figure}

\subsection{L1G2 and landmark partition algorithms}
We reiterate that Left-foot and Right-foot play dual roles, on one hand, and are comparable or even symmetric, on the other hand. Their two sets of triplet time series are highly associated. We denote the [Left-foot + Right-foot] as the L+R, for short. Thus, we will encode L+R system locally first, and then integrate L+R system with Waist or Wrist. That is, we make the L+R system a foundation to grow the integrated musculoskeletal system. For this integrative task, we develop a rather simple algorithm based ``local-first and global-second (L1G2)'' coding scheme in this section.

This L1G2 coding scheme is devised by first applying HC algorithm onto the stacked version of $ X-$, $Y-$ and $Z-$ directional time series from the Left-foot and Right-foot sensors to generate a clustering tree. Upon this tree, we pick a 10-cluster composition to form a set of 10 code-words.  Accordingly, Left-foot's triplet time series are transformed into a 1D symbolic code sequence, so is the Right-foot's. We then simply couples these two code sequences into a 2D L+R system-state trajectory. This choice of 10 code-words is supported by results of complexity evaluations in our Lampel-Ziv experiments in Fig.\ref{fig:lzcomplexity}.\\
\rule{13cm}{0.8pt}\\
\textbf{Alg.1:} Local-first \& Global-second (L1G2) Coding\\
\rule{13cm}{0.4pt}\\
\textbf{Denote:}\\
$\{(X_{L}(t), Y_{L}(t), Z_{L}(t)), \; 1\leq t \leq T\}$ from Left-foot sensor\\ $\{(X_{R}(t), Y_{R}(t), Z_{R}(t)), \; 1\leq t \leq T\}$ from Right-foot sensor\\
\textbf{(1)} Stack two time series and build a $3 \times 2T$ matrix,\\
${\cal M}_{L+R}[\cdot,1:T]=\{(X_{L}(t), Y_{L}(t), Z_{L}(t)), \; 1\leq t \leq T\}$\\
${\cal M}_{L+R}[\cdot,(T+1):2T]=\{(X_{R}(t), Y_{R}(t), Z_{R}(t)), \; 1\leq t \leq T\}$\\
\textbf{(2)} Apply HC on the temporal (column) axis of ${\cal M}_{L+R}$ to obtain $H$ clusters, coded as $\{ a_1,..., a_{H}\}$, which represent local-system states.\\
\textbf{(3)} Represent 3D time series $\{(X_{L}(t), Y_{L}(t), Z_{L}(t))$ and\\ $\{(X_{R}(t), Y_{R}(t), Z_{R}(t))$ as 1D $H$-digital time sequence $\{S_{L}(t)\}$ and $\{S_{R}(t)\}$, respectively.\\
\textbf{(4)} Couple the two local system-state time series of Left-foot and Right-foot in a 2D L+R system-state time series with 2D vector $S_{L+R}(t)=( S_{L}(t), S_{R}(t))^{'}, \; 1\leq t \leq T$.\\
\textbf{(5)} Integrate encoded Waist and encoded L+R system by a 3D $(L+R)+W$ system-state time series with 3D vector $S_{(L+R)+W}(t)=( S_{L}(t), S_{R}(t),S_{W}(t))^{'}$.\\
\rule{13cm}{0.8pt}

\begin{figure*}[h]
\centering
\includegraphics[width=5.2in]{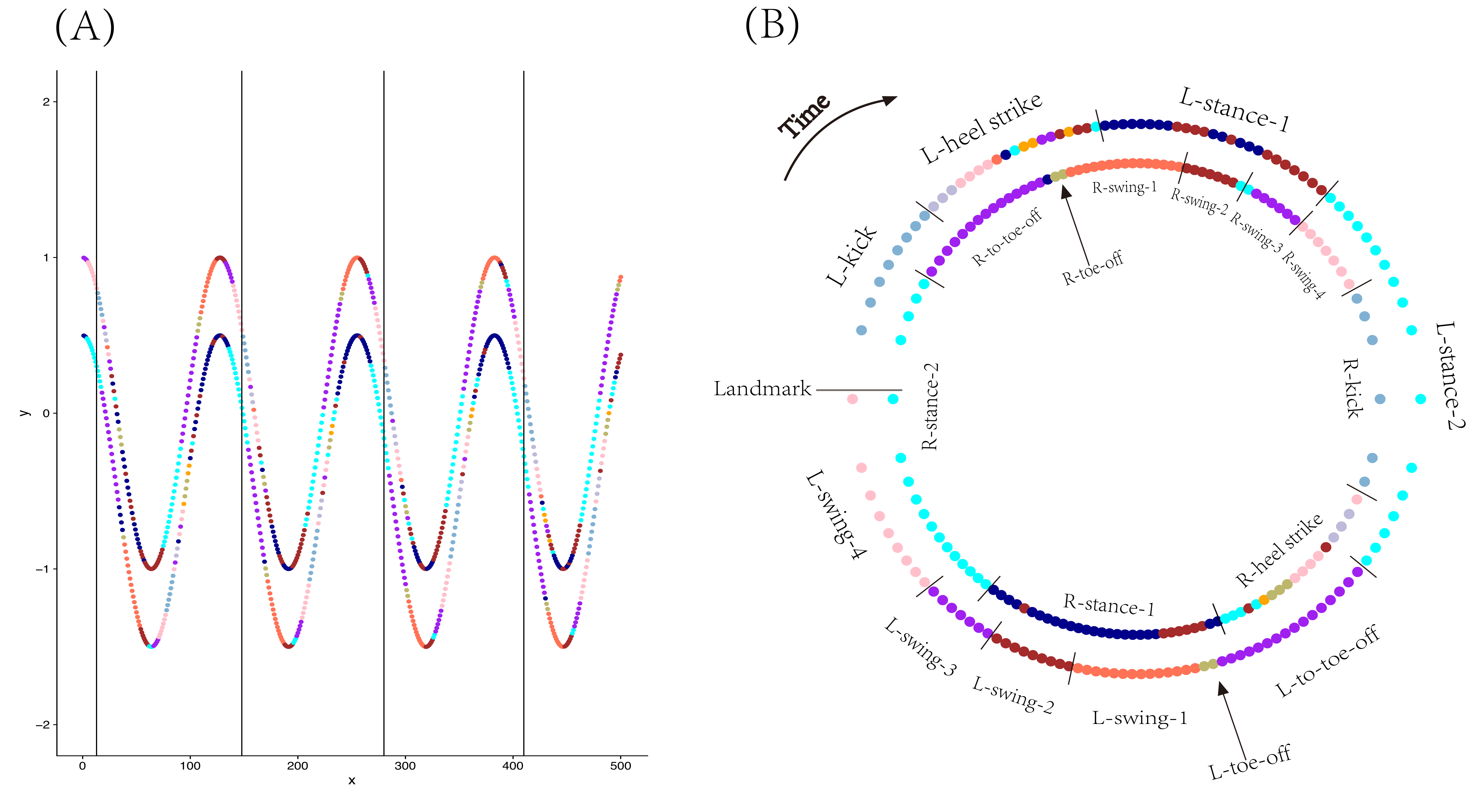}
\caption{ Color-coded rhythmical cycles in L+R system of subject $\# 5$ marked with serial biomechanical phases. (A) The coupled color coding time series on temporal period [1, 500] (Upper curve for Left-foot, Lower curve for the Right-foot. The landmarks are marked with vertical black lines; (B) Rhythmic cycle, the 3rd one in panel (A), is represented by two concentric rings (Outer ring for Left-foot, and inner right for Right-foot). 
The temporal coordinates go clockwise.}
\label{fig:colorcodes}
\end{figure*}

Next we develop a landmark algorithm to partition symbolic system-state trajectories into rhythmic cycles.\\
\noindent\rule{13cm}{0.8pt}\\
\textbf{Alg.2:} Landmark Partition\\
\rule{13cm}{0.4pt}\\
Let a $Run_i$ be a temporal segment that one specific state $i$ consecutively repeats itself.\\
Consider the 2D L+R system-state time series $\{S_{L+R}(t)\}$\\
\textbf{(1)} Calculate variance of the size of $Run_i$.\\
\textbf{(2)} Calculate variance of the recurrence time of $Run_i$.\\
\textbf{(3)} Choose the system-state $i^*$ as a ``landmark'',\\
\hspace*{15pt}$i^*= argmin_i \{ Var$(size of $Run_i$)\\ \hspace*{35pt} + $Var$(recurrence time of $Run_i$)$\}$\\
\textbf{(4)} Employ the landmark $i^*$ to partition the entire system-state trajectory into pieces of rhythmic cycles.\\
\rule{13cm}{0.8pt}

Throughout our experimental explorations across many subjects, we found that rhythms in the L+R system are rather stable, while Waist and Wrist sensors' system-state are also rhythmic, but their stability are weak. Further computed landmarks are found to coincide with the beginning of a system-state in L+R system, which is defined by a code-word pertaining to either Left-foot or Right-foot sensors, see Fig.\ref{fig:coding}. This uncertainty is likely due to some degrees of asymmetry between left foot and right foot.

\subsection{Color coded rhythmic cycles}
We apply the L1G2 algorithm onto the L+R system of subject $\# 5$ on temporal period $[1, 10,000]$. The Local coding scheme is worked out on a stacked $3\times 20,000$ matrix. The 10 code-words are color-coded, so that the identified system-states of L+R system are visible and readable with biomechanical meanings, as shown in Fig.\ref{fig:coding}.

Each colored code sequences of Left-foot and Right-foot sensors respectively achieves a dimension reduction: from 3 to 1. By coupling the two colored-codes sequences, as shown in panel (A) of Fig.\ref{fig:colorcodes},  L1G2 algorithm results cosine function like rhythm under L+R system. The symmetry on both feet are also explicit. We then apply the landmark computing algorithm on such a 2D coupled colored-code sequence on the temporal period $[1, 10,000]$ to result 77 rhythmic cycles. The average period length and standard deviation as calculated as $127.56 \pm 2.31$.

To better visualize the progressing of system-state of L+R system via coupled colored-codes, a rhythmic cycle is specifically represented by two concentric circles: Outer one for Left-foot and inner one for Right-foot, starting from the marked landmark located at the 9 o'clock position, as shown in panel (B) of Fig.\ref{fig:colorcodes}. Biomechanical phases on both feet are annotated. Indeed the gait dynamics within a rhythmic cycle is evidently revealed with deterministic and stochastic structures as characterized as follows:

\noindent{\textbf{Deterministic structures:}}\\
\textbf{A.} The process of 2D coupling-phases as its state trajectory (with clockwise temporal coordinates) is nearly deterministic throughout all computed cycles:\\ Starting from ``landmark'' $\Rightarrow$ (LF-Kick, RF-Stance2)
$\Rightarrow$
(LF-HeelStrike, RF-toToeOff)
$\Rightarrow$
(LF-HeelStrikeEnd, RF-ToeOff)
$\Rightarrow$
(LF-Stance1, RF-Swing1)
$\Rightarrow$
(LF-Stance1, RF-Swing2)
$\Rightarrow$
(LF-Stance1, RF-Swing3)
$\Rightarrow$
(LF-Stance2, RF-Swing4)
$\Rightarrow$
(LF-Stance2, RF-Kick)
$\Rightarrow$
(LF-ToeOff, RF-HealStrike)
$\Rightarrow$
(LF-ToeOff, RF-HeelStrikeEnd)
$\Rightarrow$
(LF-Swing1, RF-Stance1)
$\Rightarrow
$(LF-Swing2, RF-Stance1)
$\Rightarrow$
(LF-Swing2, RF-Stance1)
$\Rightarrow$
(LF-Swing3, RF-Stance2)
$\Rightarrow$
(LF-Swing4, RF-Stance2)
$\Rightarrow$ End at next ``landmark'';\\
\textbf{B.} A Toe-off phase of one foot has to happen after the end of Heel-strike phase of the other foot;\\
\textbf{C.} The end of kick phase as the ending phase of swing process on one foot coincide with the beginning of ``to-Toe-off'' phase.

\noindent{\textbf{Stochastic structures:}}\\
\textbf{A.} Each 2D coupling-phase varies with lengths (seen through the 3D plot of rhythmic cycles from \#3 to \#70). This is the median-scale aspect of stochasticity within a rhythmic cycle;\\
\textbf{B.} The fine-scale stochasticity is seen in the phases of ``heel-strike'' of both left foot and right foot. The variations are far from being completely random;\\
\textbf{C.} There are some orders involving with a limited number of colored nodes. The large-scale of stochasticity is seen via one or two distinct colored nodes being inserted between two phases specifically located at the two concentric circles;\\
\textbf{D.} There is also evident asymmetry on color coding of stance between the left foot and right foot.

\section{Graphic display of structural dependency in gait dynamics}
\begin{figure}[h]
\centering
   \includegraphics[width=3.3in]{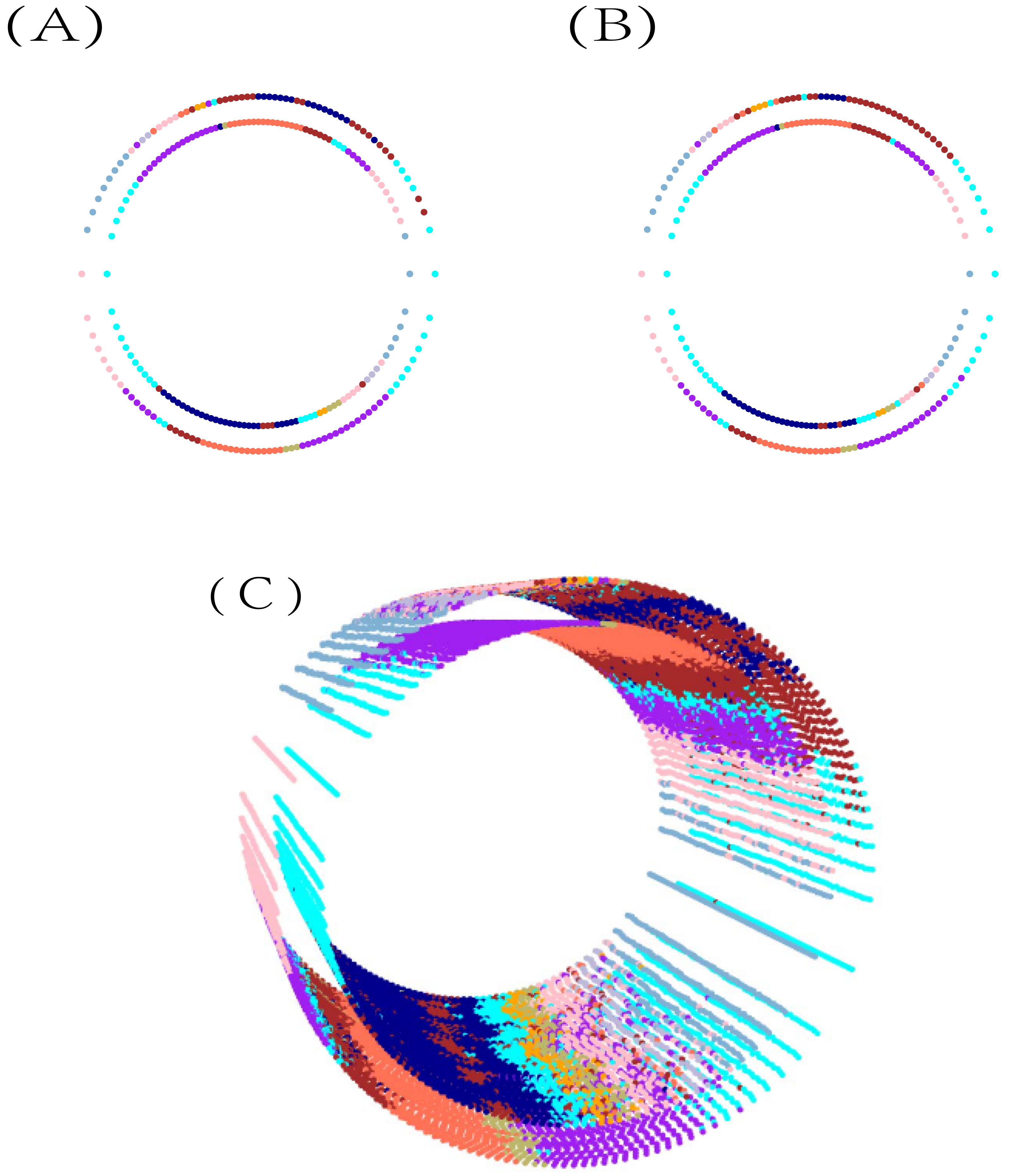}
\caption{ 3D cylinder representation of evolution of rhythmical cycles in L+R system of subject $\# 5$. (A) Concentric-ring for a rhythmic cycle from the middle of $[1, 10,000]$; (B) Concentric-ring for a rhythmic cycle from the final part of $[1, 10,000]$; (C) 3D cylinder representation of evolution of rhythmic cycles from the 3rd to the 70th.}
\label{fig:3Dcycles}
\end{figure}

The explicit deterministic and stochastic structures in panel (B) of Fig.\ref{fig:colorcodes} prescribe the structural dependency of gait dynamics in L+R system.  Such a concentric-ring representation of a rhythmic cycle within L+R system is indeed very stable. Two more rhythmic cycles: one is from the middle and another one from the end of the temporal period $[1, 10,000]$ among the 77 cycles, are rather similar,  as shown in panels (A) and (B) of Fig.\ref{fig:3Dcycles}. The great degree of stability of gait dynamics pertaining to the L+R system is also seen through a 3D cylinder representation in panel (A) of Fig.\ref{fig:3Dcycles}.

Such stability implies remarkable adaptability and precision of gait dynamics and its underlying structural dependency. The adaptability is primarily due to the interplay of deterministic and stochastic structures on the left and right foot. The deterministic structures give rise to a ``typical'' 2D coupling phase trajectory, while stochastic ones seemingly allows variations in lengths to happen among many components (or phases) of the typical cycle with total precision being about 36ms (=:4600/128). Such a precision is possible only when the deterministic structures are governed strictly by the biomechanics of human musculoskeletal system.

\subsection{Integrating Waist sensor into L+R system}
\begin{figure}[h]
\centering
   \includegraphics[width=3.7in]{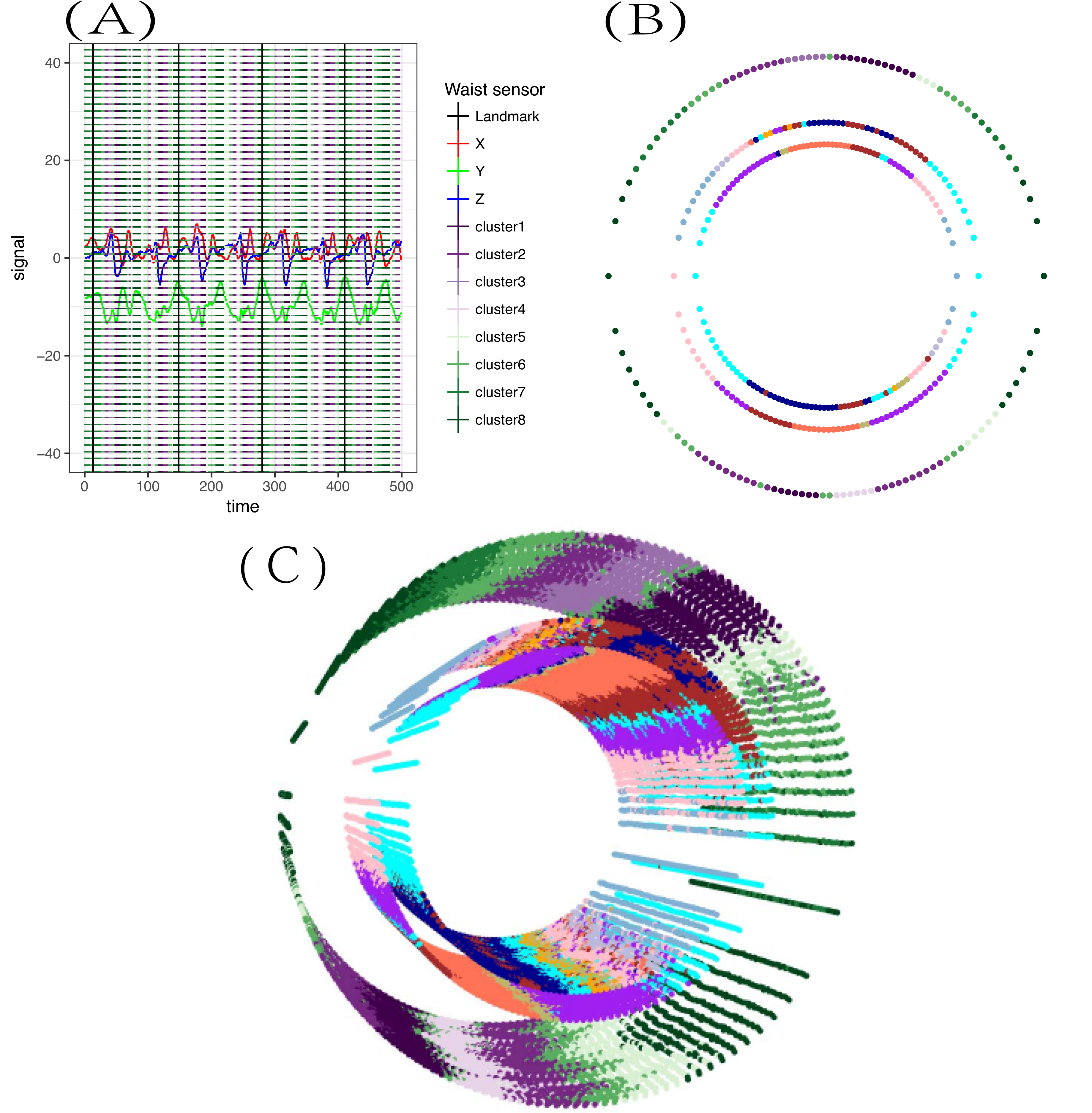}
\caption{ Integrated gait dynamics of Waist and L+R system. (A) Color coded 3D time series from waist with 8 clusters resulted from the local coding scheme of L1G2 algorithm. (B) Result of L1G2 algorithm represented by 3 layers of concentric-ring pertaining to the 3rd rhythmic cycle on the temporal period $[1, 10,000]$; (C) 3D cylinder representation of evolution of rhythmic cycles from the 3rd to the 70th of this integrated system of three sensors.}
\label{fig:3Dwaist}
\end{figure}

After constructing the rhythmic gait dynamics in L+R system, we then integrate it with the waist sensor. By applying the L1G2 algorithm on the 3D time series from Waist sensor, the resultant local coding sequence is reported in panel (A) of Fig.\ref{fig:3Dwaist}, while the results derived from the global coding scheme is reported in panel (B) of Fig.\ref{fig:3Dwaist} for one rhythmic cycle with 3 layers of concentric circles. A 3D cylinder from 3rd to 70th rhythmic cycles is built and reported in panel (C) of Fig.\ref{fig:3Dwaist}. It is clear that 3D time series from Waist sensor is rhythmic. But the rhythm is not symmetric with respect to dynamics in L+R system. Likewise, the Wrist sensor can be integrated with L+R system as well.

\begin{figure*}[h]
 \centering
   \includegraphics[width=5.4in]{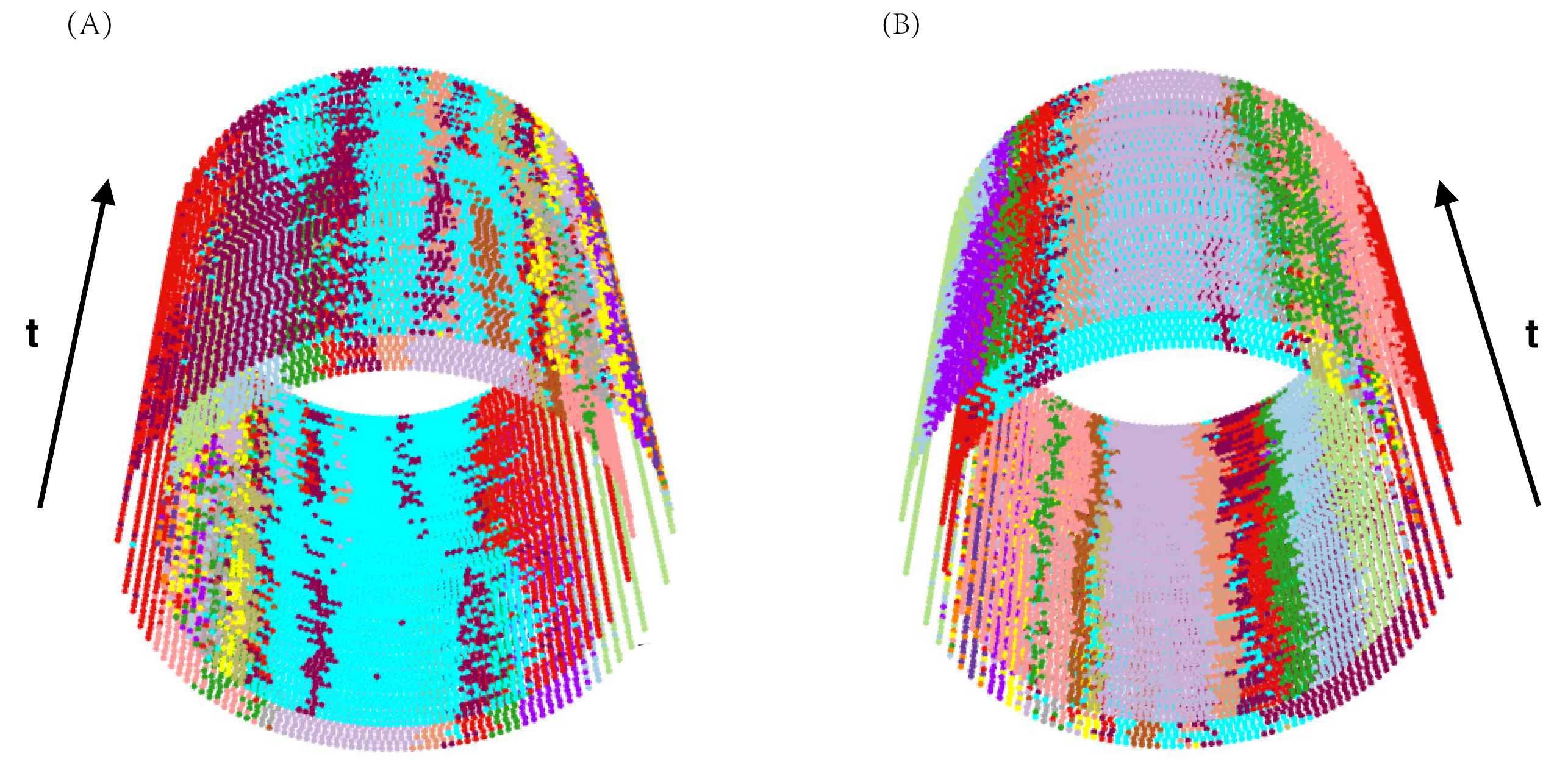}
 \caption{Two angle-views of 3D passtensor constructed from subject $\#5$'s treadmill walking with slope changes in the middle of the temporal period in $t$. The slope changes cause very subtle change on (A).}
 \label{fig:ptensorslope}
 \end{figure*}
 
\subsection{Passtensors for individual authentications}
The applications of coherently computed gait dynamics are rather wide and diverse. Here we mention two essential one in passing without going into details, and then focus on cybersecurity. The first comment is that this L1G2 algorithm will allow us to integrate acceleration sensors with gyroscope sensors. By combining the two kinds of sensors, the resultant gait dynamic system will be rather complex, but extremely interesting. The second comment is obvious that such a 3D representation can be utilized as a platform for mimicking the entire gait dynamics captured by time series data derived from the four acceleration sensors. Such a task of building realistic mimicry of a complex system is technically very challenging, while is scientifically very important, for instance in robotics. Up to now, robots still walk in very unhuman-like fashions. This issue might be resolved to great extent by incorporating gait dynamics.

Now we turn to cybersecurity, clinical diagnosis and self-evaluating individual health statuses. It becomes clear that, based on our 3D graphic displays of gait dynamics, an individual's process of rhythmic cycle is characterized by the evolution of cyclic deterministic phases with individual specific twists as well as idiosyncratic stochastic deviations associated with all phases. Hence, a 3D cylinder indeed becomes a basis for authenticating this particular individual. For this use, such a 3D cylinder is called ``passtensor''. More specifically speaking, a L+R system's deterministic cycle of 2D biomechanical phases: from one landmark proceeding to the next one, in indeed provides a rigid frame, while the stochastic phases' lengths and presence or absence of some color codes between adjacent phases provide the soft frames for the purposes of authentications. This authentication capacity further illustrated as follows. For instance, consider the subject $\#5$ in MAREA walked on a treadmill with slope change: from horizontal($0^{\circ}$) to $5^{\circ}$ during a recording period. This person's 3D passtensor corresponding to this period is shown in Fig.\ref{fig:ptensorslope} with two views from two different angles. The angle specific view in panel (A) of Fig.\ref{fig:ptensorslope} reveals visible changes. Such changes are likely critical patterns for authentication purposes.

Here we briefly reiterate the practical uses of our 3D cylinder graphic display of gait dynamics in self-evaluating individual health statuses. By stacking two temporal segments of gait time series from two different temporal periods, we can examine the degrees and aspects of similarity and differences in regarding to deterministic and stochastic structures between these two temporal segments. This is an effective way of finding out subtle and minute discrepancies to serve the early warning purposes.

\section{Conclusion in system complexity}
Our first theme of data-driven computing paradigm, PSSA, allow us to include many principle gait states as a collective of key characteristics for identifying as many people as we want. From many aspects, this identification approach is indeed very distinct from identifications based on facial and voice recognitions, finger-print or retina scanning. It is much easier to achieve social unbiasedness. It is much more difficult to imitate or to fake. 

Our second theme of data-driven computing paradigm, consisting of L1G2 coding and landmark algorithms, enables us to explicitly manifest multiscale dynamic patterns of gait dynamics. The graphic displays of single rhythmic cycle and collective 3D passtensor clearly demonstrate how the deterministic circle of biomechanical phase couples with stochastic variations sprinkling between consecutive phases, and offer a whole-view of an individual's gait dynamics. Such intricate coupling relations between deterministic and stochastic structures are the backbones of structural dependency of gait dynamics. They retain essential basis for mimicking an individual's gait dynamics in animation. Its practical uses in clinical diagnosis and cybersecurity are also evident. In fact, the original motivations of this gait study is aiming at detecting relative minor changes in gait dynamics for healthy peoples and gesture tuning for athletes. These two topics require very detailed structures within personal dynamics.

From computational science perspective, our PSSA and L1G2 coding algorithm rests on the crucial fact: different time series have different functions linking to different subsystems of a complex system of interest, so they should not be treated equally and uniformly. Such a rationale is a key for revelations of multiscale structural dependency. It is also the key rationale for recreating a system's authentic dynamics.  Overall, good design of graphic displays definitely pave avenues for true understanding onto a complex system.


\newpage

\appendix
\section{Supplementary}

\subsection{Data Source} The MAREA Gait Database is available at: \url{http://islab.hh.se/mediawiki/Gait_database}. The Human Gait Database(HuGaDB) is available at: \url{https://github.com/romanchereshnev/HuGaDB}.


\subsection{Coverage proportion curves of Principle System-States}
\begin{figure}[h]
\centering
\includegraphics[width=3in]{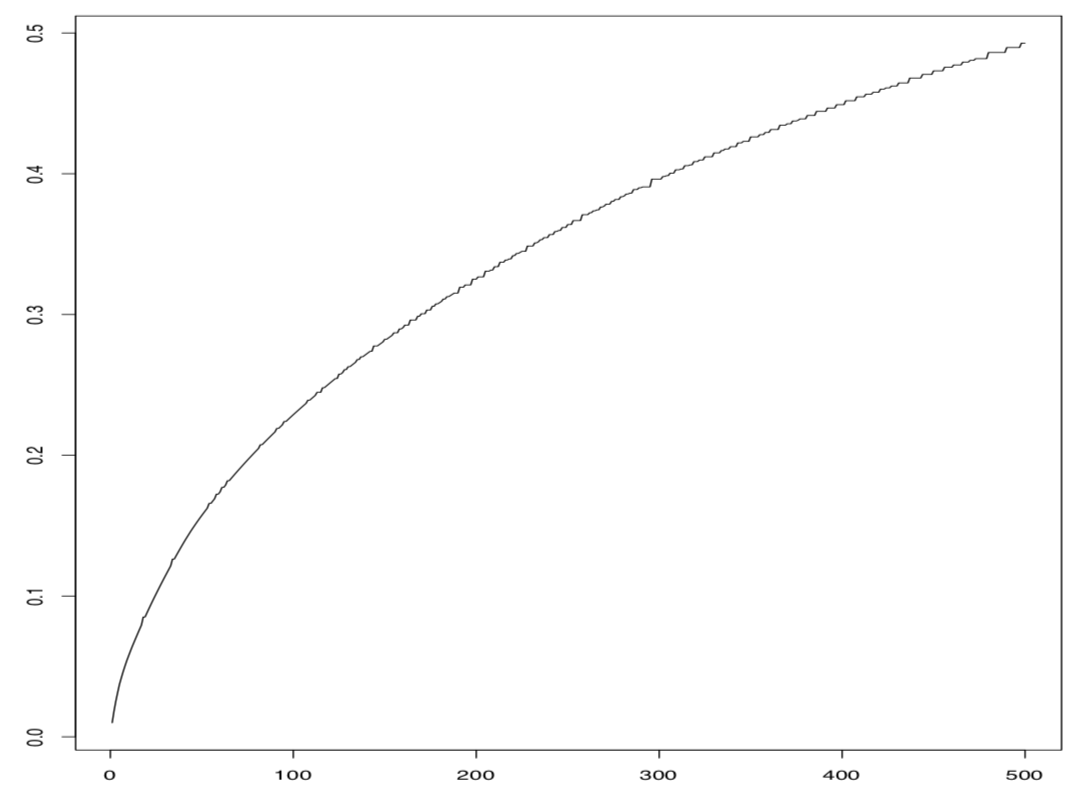}
\caption{$(r(N^*) \, v.s \, N^*)$ based on 9-dim gait time series from 3 sensors fixed at Left foot and Right foot and wrist among 10 subjects in MAREA database. The triple coding is based on $\alpha=0.3$ and $\beta=0.7$ quantiles. }
\label{fig:PSS_CurveA}
\end{figure}

\begin{figure}[h]
\centering
\includegraphics[width=3.3in]{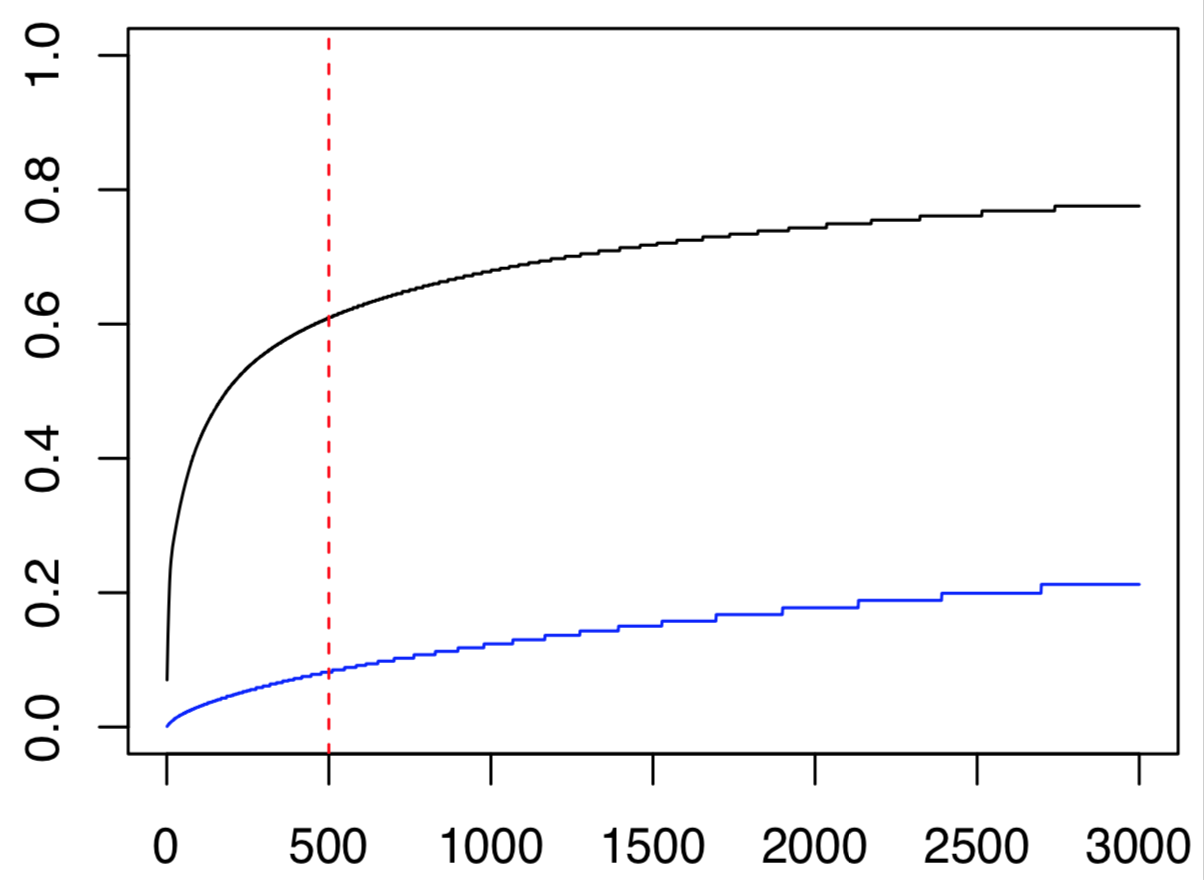}
\caption{$(r(N^*) \, v.s \, N^*)$based on 18-dim gait time series derived from 6 sensors fixed to left and right thighs, shines and feet in HuGaDB database. The black curve is pertaining to the triple coding based on $\alpha=0.1$ and $\beta=0.9$ quantiles, while the blue curve is based on $\alpha=0.3$ and $\beta=0.7$ quantiles}
\label{fig:PSS_CurveB}
\end{figure}

\begin{figure*}[h!]
 \centering
   \includegraphics[width=4.5in]{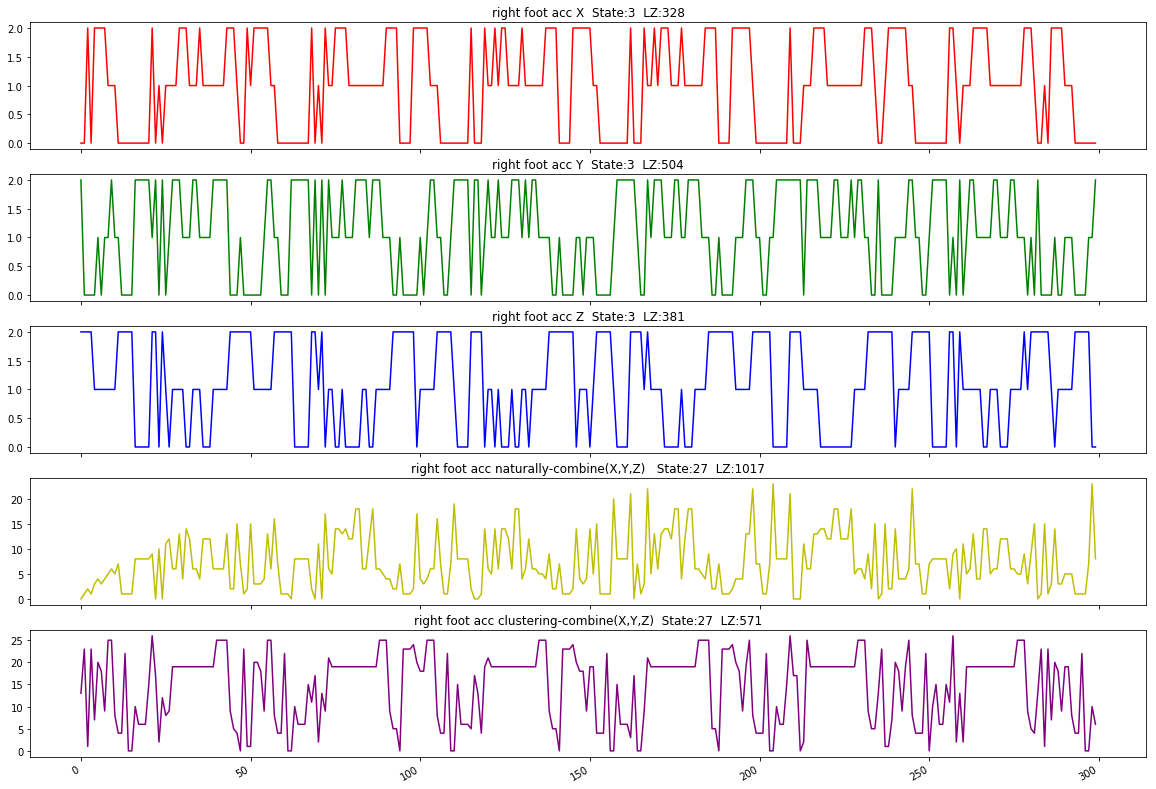}
 \caption{From top to bottom, code each accelerometer time series from \textbf{right foot} separately and combine them into one sequence in two different ways; one is a natural way of combination (the second last to the bottom), the other is our clustering-based combination (the last); ours has less LZ complexity around a half compared with the natural way; the seasonal pattern is shown clearer and more rhythmic in our method}
 \label{fig:compare_rf}
 \end{figure*}

\begin{figure*}[h!]
 \centering
   \includegraphics[width=4.5in]{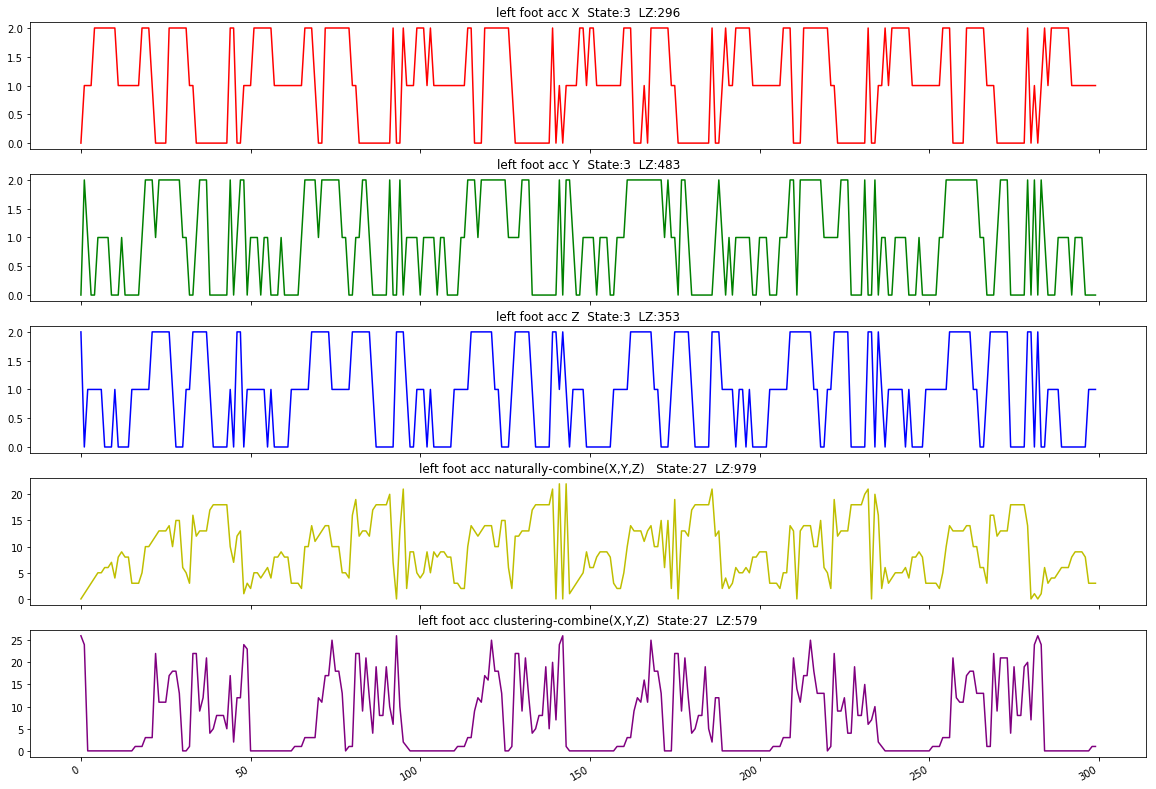}
 \caption{From top to bottom, code each accelerometer time series from \textbf{left foot} separately and combine them into one sequence in two different ways; one is a natural way of combination (the second last to the bottom), the other is our clustering-based combination (the last); ours has less LZ complexity around a half compared with the natural way; the seasonal pattern is shown clearer and more rhythmic in our method}
 \label{fig:compare_lf}
 \end{figure*}

\end{document}